\documentclass[useAMS,usenatbib]{mn2e_astroph}

\usepackage{graphicx}

\def\nct#1{\nocite{#1}}

\title[The geometry of a radio pulsar beam]
{The geometry of a radio pulsar beam
}

\author[J.~Dyks]
{J.~Dyks
\\
Nicolaus Copernicus Astronomical Center, Rabia\'nska 8, 87-100, Toru\'n,
Poland\\
}
\begin{document}

\date{Accepted .... Received ...; in original form 2017 April 10}


\maketitle

\label{firstpage}

\begin{abstract}
Taxonomy of radio pulsar profiles  
is mostly based on a system of Ptolemaic artificiality, consisting of 
separated rings and a core, 
arbitrarily located at disparate altitudes in 
the magnetosphere. 
Diversity of observed profile shapes 
 clearly exceeds
the interpretive capability of such conal model. 
Moreover, bifurcated features observed 
in pulsar profiles 
 imply a system 
of fan beams radially extending away from the dipole axis. 
The bifurcations 
can be understood as the imprint of the elementary radiation pattern of
the long-sought radio emission mechanism, thus identifying the 
latter. 
Their size, 
however, 
is several times larger than implied by the curvature of magnetic dipole 
lines.
Here I show 
that the illusion of disconnected rings and the size of bifurcated features 
can be explained through a natural geometry which combines 
the properties of both 
the cone and the fan beam. It is a flaring spiral which makes several 
revolutions around the dipole axis on its way to leave the magnetosphere.
Such geometry 
is consistent with a stream of outflowing and 
laterally drifting plasma. The bifurcated components are so
wide, because the curvature on such a spiral is larger  
than that of the dipolar magnetic field, hence they are consistent 
with the extraordinary mode curvature radiation. 
\end{abstract}


\def\lap{\hbox{\hspace{4.3mm}}
         \raise1.5pt \vbox{\moveleft9pt\hbox{$<$}}
         \lower1.5pt \vbox{\moveleft9pt\hbox{$\sim$ }}
         \hbox{\hskip 0.02mm}}

\def\rwobs{R_W}
\def\rwcon{R_W}
\def\rwstr{R_W}
\def\winobs{W_{\rm in}}
\def\woutobs{W_{\rm out}}
\def\phm{\phi_m}
\def\phmi{\phi_{m, i}}
\def\thm{\theta_m}
\def\dres{\Delta\phi_{\rm res}}
\def\win{W_{\rm in}}
\def\wout{W_{\rm out}}
\def\rin{\rho_{\rm in}}
\def\rout{\rho_{\rm out}}
\def\phin{\phi_{\rm in}}
\def\phout{\phi_{\rm out}}
\def\xin{x_{\rm in}}
\def\xout{x_{\rm out}}

\def\thmin{\theta_{\rm min}^{\thinspace m}}
\def\thmax{\theta_{\rm max}^{\thinspace m}}

\section{Introduction}
\label{intro}

Since the discovery of pulsars in 1967 (Hewish et al.~1968)
\nct{hbp68}
thousands of pulse profiles have been observed at different radio frequencies 
$\nu$ (Hankins \& Rankin 2010, hereafter HR10; Mitra et al.~2015, hereafter
MAR15; Dai et al.~2015).
\nct{hr10, mar2015, dhm15} 
 Some of the profiles are approximately symmetric, which 
has led to the nested cone model of the radio emission beam
(Ruderman \& Sutherland 1975; Backer 1976; Rankin 1983) 
\nct{bac76, rs75, ran83} 
-- the main model in use so far. 
The corresponding emission region consists of two rings 
and a low-altitude filled-in 
core region, all centered at the dipole axis 
and localised at well separated altitudes in
pulsar magnetosphere (Rankin 1990, 1993; Gangadhara and Gupta 2001;
although compare Wright 2003).
\nct{ran90, ran93, gg01, wri03} 
The model suffers from 
multiple problems: the selection of up to 
three altitudes lacks physical justification 
and offers limited understanding of the large diversity 
of profile shapes. The profiles are often highly asymmetric and have 
components with flux ratio which curiously evolves with frequency.
 The latter 
effect has led to the idea of general patchiness of the beam
 (Lyne \& Manchester 1988; Karastergiou \& Johnston 2007).
\nct{lm88, kj07}
The flux ratio reversal in components observed at different $\nu$ 
is a ubiquitous phenomenon, observed even for profiles that look 
highly symmetric at some  $\nu$ (e.g.~PSR B0525$+$21,
B0301$+$19, B1133$+$16, HR10, MAR15).
\nct{hr10, mar2015}
In other profiles the change of flux ratio with $\nu$ 
has extreme magnitude,  
with components
disappearing at some pulse longitudes, and appearing at others 
(e.g.~PSR B1541$+$09, B1737$+$13, B1821$+$05, HR10).
\nct{hr10}
Moreover, 
precursor and postcursor components appear on either side 
of
 many profiles as additional features. Some of them, 
but also the usual ``conal" components
(e.g.~B1933$+$16, Mitra et al.~2016; 
\nct{mra2016} B1946$+$35, Mitra \& Rankin 2017) \nct{mr2017}
 belong to the class of bifurcated features 
 (split emission components and double notches). These are often
 observed far from the main pulse, and may have 
 highly symmetric, double form that has been attributed 
directly 
to the elementary radiation pattern of the curvature radiation 
(PSR J1012$+$5307, Dyks et al.~2010, hereafter DRD10; 
\nct{drd10}
J0437$-$4715, Navarro et al.~1997; Os{\l}owski et al.~2014).
\nct{nms97, ovb14}

\section{The new emission geometry}

\begin{figure}
\includegraphics[width=0.48\textwidth]{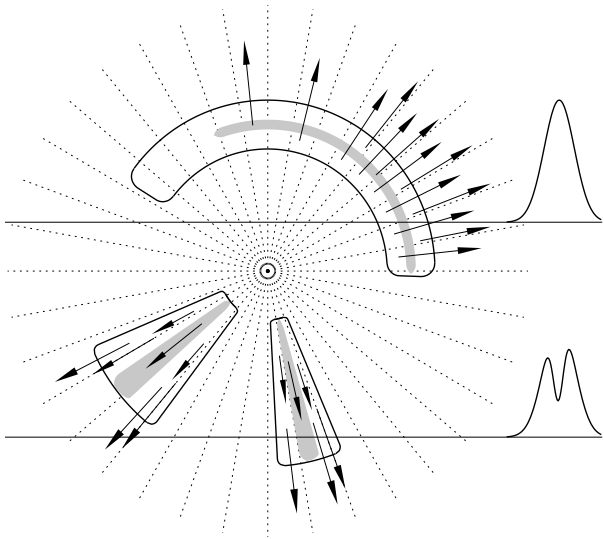}
\caption{
A face on view of two pulsar beams: the conal beam (top)
and the fan beam (bottom), illustrating the conundrum of  
bifurcated components.
The flux minimum at the center of a BC requires a lowered emissivity 
at the center of each beam, as marked with the grey bands. 
In the case of the conal model, the minimum should be smeared out by
 the radial plasma outflow, marked with the arrows. For the fan beam 
geometry, the flow does not destroy the feature. Dotted lines present 
the sky-projected $B$-field lines.
}
\label{bif}
\end{figure}

As explained in Fig.~1, 
the nested cone model is helpless with the bifurcated components (BCs), since 
the flaring shape of dipolar magnetic field lines  
implies the outflow 
of the radio emitting plasma away from the dipole axis. 
To produce 
the central dip in a BC, the charges would have to stop
emitting for a short while, and right at the moment 
when they are crossing the middle of the emission region.
To explain the BCs, 
the fan beam geometry has therefore been rediscovered (DRD10),
\nct{drd10} 
after years of neglect since 1987, when it was first 
suggested (Michel 1987). \nct{m87} 
In the fan beam model, 
the beam 
extends along the trajectory of plasma motion (Fig.~1) so
the presumed radial motion of plasma does not smear out the bifurcation.

However, there exists a geometry which is natural from 
the point of view of physics, and combines the geometric 
properties of a cone and a fan beam. 
It is a flaring spiral (or flaring helix) which makes several
revolutions around the dipole axis while the plasma is streaming along this 
helix towards the light cylinder.
When the observer's sightline is traversing through coils of such a spiral, 
 pairs of altitudes are detected, which decrease towards the 
profile center. This creates the misleading illusion of nested cones, 
which has  
been the rule in taxonomical identification of profiles for decades. 

\begin{figure}
\includegraphics[width=0.48\textwidth]{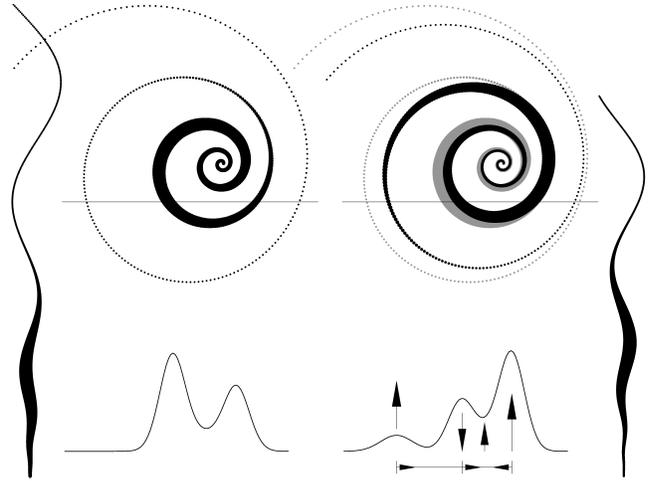}
\caption{A spiral radio pulsar beam viewed head on 
at two different frequencies (the left low-$\nu$ spiral 
is plotted on the right 
in grey for reference). Side views of the associated emission region are shown near the
left and right edge of the figure. 
The local radio emissivity along each spiral is marked by
line thickness, and 
reaches maximum at the altitude of $3\%$ (left) and $4\%$ (right) of the light
cylinder radius. 
The thick parts of the spirals
are therefore misaligned in 
azimuth. The horizontal straight sections mark the path 
of sightline through the beams, and result in the profile 
differences shown 
below. Note the seemingly random 
change of the profile with $\nu$. 
The origin of each spiral is dislocated rightwards, because of the
AR effects, which in the presented case  
are smaller at higher $\nu$ 
despite the higher locus 
of the average high-$\nu$ region. 
Both spirals are anchored at the footpoint parameter of $0.22$
and assume $|\vec v|\approx c$, $v_\phi=Ac(r/R_{ns})^k$, and 
 $k=0.75$. For the left spiral $A=5\cdot10^{-3}$ whereas
 $A=5.5\cdot10^{-3}$ on the right.
}
\label{spiral}
\end{figure}

Assuming that radio emission observed at different frequencies $\nu$ 
is on average 
generated at different altitudes (or radial distances $r$), different 
segments of the spiral, located further from, or closer to its origin 
are radio bright (detectable). 
With the change of $\nu$, the brightest part of emission
region is then rotated in the magnetic azimuth, 
as measured around the dipole axis.
The brightest part of the ensuing beam is shown in Fig.~2 as the increased 
thickness of the spiral which is viewed face on.
As can be seen in the figure, the nontrivial evolution of the 
components' flux ratio with $\nu$
is an
inherent feature of the flaring spiral beam. Similarly, the disappearance 
of components, and their appearance at a new pulse longitude occurs easily.
Thus, the prime result of the $\nu$-dependent emission altitude is 
 the $\nu$-dependent components' flux ratio.

The profiles are also well known of widening at low
frequencies (Mitra \& Rankin 2002),
\nct{mr02} 
i.e.~the peak-to-peak separation between 
components increases. 
This implies that 
at higher $\nu$ the sky-projected spiral must be coiled more tightly, 
i.e.~more energetic electrons must follow a helix that 
is anchored or revolving 
closer to the dipole axis. 
The radio waves,
emitted mostly tangentially to the helical trajectory, 
are emitted at the opening angle 
$\theta_s(r)\approx \arctan(v_\phi/v_t)\approx v_\phi/v_t$, 
where $v_\phi$ and $v_t$ 
are the azimuthal and tangent-to-$B$ velocity components. 
I attribute the spiral geometry to the $\vec E\times \vec B$ drift
\emph{within} the radio emission region, where $\vec E=\vec E_t + \vec
E_\perp$ and $E_t$, 
$E_\perp$ are the electric field components parallel and orthogonal to the
local $B$. 
When only the magnitude 
of $\vec E$ is fluctuating or nonuniform, $E_\perp$ and $E_t$ are 
 positively correlated. The strong $E$ induces both
 the large Lorentz factor $\gamma$ 
(hence high $\nu$) 
and the large $v_\phi$ 
which enables the electrons to 
make the inner coils at lower altitudes, where the $B$ flux tube 
(and the resulting spiral) is narrower. As verified numerically, 
for a larger $v_\phi$ (keeping total $v\approx c$), the coils 
are produced at smaller $r$ and the spiral is 
contracted.\footnote{This is not inconsistent with the formula
$\theta_s\approx v_\phi/v_t$, since the latter is valid for a fixed $r$.}
The spread of 
$E$ then provides a natural single cause for the association of  
high $\nu$, low $r$ and small $\theta_s$. However, when the direction of
$\vec E$ is fluctuating (instead of the magnitude), then 
$E_t$ and $E_\perp$ (hence $\nu$ and $v_\phi$) 
are anticorrelated and the high $\nu$ emission may form a wider spiral at 
larger $r$. 
With the unscreened $\vec E$ within the radio emission region, 
the continuing acceleration increases the electrons' energy 
with $r$, as long as the coherent radiative losses are smaller than the energy
gain. This also implies the increase of emitted $\nu$ with $r$.
Note that the larger average altitude of the high-$\nu$ emission 
can be consistent with smaller aberration-retardation (AR) effects
when $v_\phi$ is larger and the detectable spiral coils are produced 
at smaller $r$ (cf.~the grey and black spirals in Fig.~2). 

Taking the above-described effects together,
it becomes possible to interpret  
instances of profile evolution which have been unaccessible 
for the nested cone model.
Fig.~3 presents the interpretation of profile evolution for 
B1541$+$09 (cf.~Fig.~8 in HR10). The core component at 
$100$ MHz becomes surrounded by two asymmetric peripheric 
components (PCs) 
at $400$ MHz.  Above 1.4 GHz the leading PC approaches 
the core and they merge with a bridge of emission in between.
Fig.~4 presents a different viewing geometry for B1821$+$05 (cf.~Fig.~9 in 
HR10). This time it is the trailing PC that 
merges with the core at high $\nu$, because of the narrowing 
of the spiral and the AR-related distortions.

\begin{figure}
\includegraphics[width=0.48\textwidth]{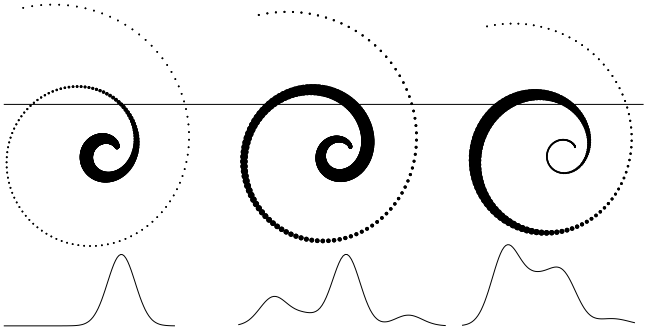}
\caption{Interpretation of the $\nu$-dependent profile evolution for 
PSR B1541$+$09. The single component at $100$ MHz (left) becomes surrounded by 
two asymmetric low flux components at $400$ MHz (middle). Further 
displacement of emission away from the spiral origin, and the contraction
of the entire spiral, make the leading component brightest and bridged 
with the central component at $1.4$ GHz (right). Cf.~Fig.~8 in HR10. 
}
\label{evol}
\end{figure}

\begin{figure}
\includegraphics[width=0.48\textwidth]{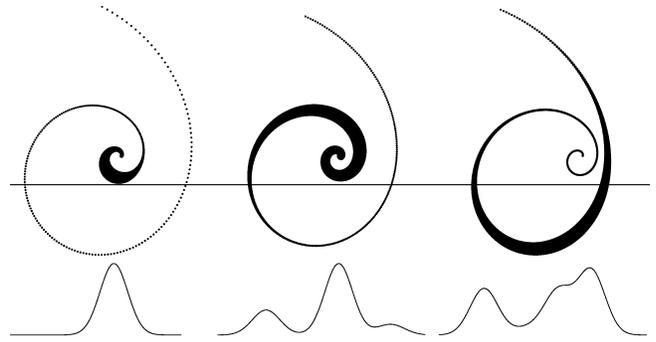}
\caption{Interpretation of the $\nu$-dependent profile evolution for 
PSR B1821$+$05. The spirals represent the sky-projected pulsar radio 
beam at $0.1$, $0.4$ and $2.4$ GHz (left to right, cf.~Fig.~9 
in HR10). The high-$\nu$ spirals are assumed to follow narrower
$B$-flux tubes and $v_\phi$ decreases with $\nu$. 
 The coils of right spirals are therefore located 
at a larger $r$ and are more distorted by the AR
effects. 
}
\label{evol2}
\end{figure}

\section{Bifurcated components and the mechanism of radio emission}
 
The bifurcated emission components (BCs), when not blurred or 
overlaid on 
other components, have highly symmetric
shape, which is very close to the elementary radiation pattern of curvature
radiation (CR) propagating in the extraordinary polarisation 
mode (DRD10; Gil et al.~2004).
\nct{drd10, glm04} 
Moreover, the peaks of the BCs approach each other with increasing $\nu$, 
at the rate consistent with the CR mechanism (Dyks \& Rudak 2012).
\nct{dr12}
However, the observed
scale of wide BCs ($\Delta_{obs}$ between $3^\circ$ and $8^\circ$) 
was an order of magnitude 
larger than the opening angle of the CR beam in dipolar magnetosphere: 
$\Delta \approx0.8^\circ f_{an}(\nu_9\rho_7)^{-1/3}$, 
where  $\rho=\rho_7 10^7$ cm is the 
curvature radius of the emitting electrons' trajectory, $f_{an}$ is 
a viewing angle factor of the order of a few, 
and $\nu_9$ is the observation
frequency in GHz. 
Since $\nu_9\sim1$ is fixed by the choice of the telescope's receiver, 
for $f_{an}\sim4$ the observed $\Delta_{obs}$ 
could be explained by the curvature radii of 
$\sim\negthinspace10^5$ cm,
 i.e.~about 
 two orders of magnitude 
smaller than radii of dipolar field lines along
which the emitting plasma was assumed to move. 
This problem is solved by the flaring helix, 
since the radius of curvature in the 
 helix can be much smaller than that of the $B$-field lines, which only 
provide the outer envelope for the helix
(see the side views of flaring helices in Fig.~2).
This same effect simultaneously explains why the widest 
BCs are observed 
at longitudes far away from the main pulse: 
the radio waves
are emitted at the angle $\theta_s\approx v_\phi/v_t$ 
which can be larger than the opening angle of the polar tube envelope. 
For a standard (cylindrical) helix with very elongated coils 
($r_\perp \ll h$, where $r_\perp$ 
is the transverse circulation radius and $h=2\pi r_\perp v_t/v_\phi$ 
-- the upward 
advance of the helix after one coil) 
the radius of curvature is $\rho
\approx h^2/(4\pi^2r_\perp)=
\theta_s^{-2}r_\perp$. 
The large
$v_\phi/v_t$ 
implies both smaller curvature radii
$\rho$ and more slanted emission direction, which is in line with 
the peripheric occurence of BCs (as precursors or PCs). 
With the smaller $\rho$ offered by the helix, the extraordinary mode CR
provides an ideal origin for the astonishing symmetry and the $\nu$-dependent 
merging of BCs. 
The formula for $\Delta$ could in principle
 be used to estimate 
the curvature radius in detectable parts of the helix. However, 
the resulting formula: $\rho_7=\nu_9^{-1}f_{an}^3(\Delta_{obs}/0.8^\circ)^{-3}$
depends in a high power on the unknown $f_{an}$ which can be anywhere
between $1$ and $\sim\negthinspace10$ (and likely larger than $1$ 
since it is roughly an inversed
product of two sine functions). 

\section{Discussion}

The exact shape of the flaring helix, 
which represents the {\it average} radio emission region, 
 depends on the unknown distribution of $\vec E$ with altitude. 
The entire helix may be fixed 
in azimuth at some surface irregularity of $B$-field or move
azimuthally in such a way that some orientation is preferred whenever 
it is crossed by the line of sight.
Single 
pulse phenomena such as different drift modes  
may suggest time variability of the entire helix. 
The radio-bright part of the helix does not
seem to extend for many azimuthal turns, because profiles with numerous 
components (of M and Q type) are rarely observed. 
The M and Q profiles tend to contain several approximately equidistant 
components which suggests that tangents to
the flaring helix may sometimes form a roughly Archimedean spiral on the sky.
Performed calculations reveal that large portion of parameter space
results in  Archimedean-like spirals, 
e.g.~they appear when 
$|\vec v| 
\approx c$, $v_\phi=A c (r/R_{ns})^k$, $A\ll1$ and $k$ is  
in the large range between
$\sim\negthinspace0.5$ and
$\sim\negthinspace10$.
The calculations also suggest that the helix is anchored near the 
dipole axis, not at the polar cap rim, as one may expect if the spacious
 polar regions of millisecond pulsars are to be filled in with emission. 
At some altitude the sky-projected spiral 
readily unwinds into a real fan beam, 
either because of the flaring of the dipolar field lines, or because 
of the value of $v_\phi/v_t$. 
The outcome is then similar to 
 the mapped beam of PSR J1906$+$0746 (Desvignes et al.~2013).
\nct{dkc13}

The flaring spiral beam solves the problem of 
 disconnected emission rings, explains the 
``approximately conal" symmetry of profiles, 
allows for the peculiar $\nu$-evolution of profiles,
 and for the existence of wide 
bifurcated components. The spiral beam opens new possibilities to interpret 
the shapes of average pulsar profiles and of the single pulse phenomena 
(such as drifting, nulling or profile moding).

\section*{acknowledgements}
I thank B.~Rudak 
and 
A. Frankowski for discussions and comments on the manuscript. 
This work was funded by 
the National Science Centre grant DEC-2011/02/A/ST9/00256.

\bibliographystyle{mn2e}
\bibliography{listofrefs}


\end{document}